\documentclass[useAMS,usenatbib]{mn2e}

\usepackage{graphicx}
\usepackage{epsfig}
\usepackage{times}
\usepackage{natbib}

\def\bm#1{{\hbox{\boldmath $#1$\unboldmath}}}
\def\vec#1{\bm{#1}}

\newcommand{\be}{\begin{equation}}
\newcommand{\ee}{\end{equation}}
\newcommand{\bea}{\begin{eqnarray}}
\newcommand{\eea}{\end{eqnarray}}
\newcommand{\hm}{{\mathcal{P}}}
\newcommand{\CMB}{{\rm{CMB}}}
\newcommand{\ISW}{{\rm{ISW}}}
\newcommand{\prim}{{\rm{prim}}}
\newcommand{\tot}{{\rm{tot}}}
\newcommand{\all}{{\rm{all}}}
\newcommand{\proj}{{\rm{\,proj}}}
\newcommand{\cc}{{cc}}
\newcommand{\rec}{{\rm{rec}}}
\newcommand{\crit}{{\rm{crit}}}

\newcommand{\ls}{{\rm{ls}}}
\newcommand{\pix}{{\rm{pix}}}
\newcommand{\av}{{\rm{av}}}
\newcommand{\loc}{{\rm{loc}}}
\newcommand{\Max}{{\rm{max}}}
\newcommand{\Det}{{\rm{det}}}

\newcommand{\g}{{G}}
\newcommand{\growth}{{\mathcal{D}}}

\title[Optimal ISW detection and joint likelihood] 
{Optimal integrated Sachs-Wolfe detection and joint likelihood for
  cosmological parameter estimation} 
\author[M. Frommert$^{1}$, T.~A. En{\ss}lin, \& F.~S. Kitaura]
{M. Frommert, T.~A. En{\ss}lin, \& F.~S. Kitaura\\
Max-Planck-Institut f\"ur Astrophysik,
Karl-Schwarzschild-Stra{\ss}e 1, D-85748 Garching b. M\"unchen, Germany\\
mona@mpa-garching.mpg.de\\
}

\begin{document}

\date{Accepted 2008 September 12. Received 2008 September 12; in
  original form 2008 July 2}
\pagerange{\pageref{firstpage}--\pageref{lastpage}} \pubyear{2008}
\maketitle
\label{firstpage}

\begin{abstract}
We analyse the local variance effect in the standard method for
detecting the integrated Sachs-Wolfe effect (ISW) via
cross-correlating the cosmic microwave background (CMB) with the large-scale
structure (LSS). Local variance is defined as the systematic noise in the 
ISW detection that originates in the realization of the matter
distribution in the observed Universe. We show that the local variance
contributes about 11 per cent to the total variance in the standard
method, if a perfect and complete LSS survey up to $z \approx 2$ is assumed.
Due to local variance, the estimated detection significance and
cosmological parameter constraints in the standard method are
biased. In this work, we present an optimal 
method of how to reduce the local variance effect in the
ISW detection by working conditional on the LSS data. The variance of
the optimal method, and hence the
signal-to-noise ratio, depends on the actual
realization of the matter distribution in the observed Universe. We
show that for an ideal galaxy survey, the average signal-to-noise
ratio is
enhanced by about 7 per cent in the optimal method, as compared to the
standard method. 
In the framework of our method, it is straightforward to
correct for the magnification bias coming from gravitational lensing effects.
Furthermore there is no need to estimate the
covariance matrix by Monte Carlo simulations as in the standard
method, which saves time and increases the accuracy.
Finally, we derive the correct joint likelihood
function for cosmological parameters given CMB  and LSS data within the
linear LSS formation regime, which
includes a small coupling of the two datasets due to the ISW effect.

\end{abstract}

\begin{keywords}
Cosmology: CMB -- Large-Scale Structure -- 
cosmological parameter estimation 
\end{keywords}


\section{Introduction} \label{intro}

The integrated Sachs-Wolfe (ISW) effect \citep{isw} is an
important probe of the existence and nature of dark energy
\citep{crittenden_turok} and the nature of 
gravity \citep{lue_gravity,zhang_gravity}. Spatial curvature also
gives rise to an ISW effect, but is well constrained to be very close
to zero by cosmic microwave background (CMB) experiments such as the
{\it Wilkinson Microwave Anisotropy Probe} \citep{wmap_5}. However, the
detection of the ISW signal remains challenging, for it is obscured by
primordial fluctuations in the CMB.
In recent years, substantial effort has been made to detect the
ISW effect via cross-correlation of the CMB 
with large-scale structure (LSS)
surveys, such as optical galaxy and quasar surveys (Sloan Digital Sky Survey,
\cite{sdss_6}, and Two-Micron All-Sky Survey, \cite{2mass}),  
radio surveys (NRAO VLA Sky Survey, \cite{nvss}), and X-ray surveys (High Energy
Astrophysics Observatory, \cite{heao}). Such 
cross-correlation
studies have, for example, been done by
\cite{x_ray_boughn}, \cite{boughn_crittenden_nature}, \cite{
boughn_crittenden_detection}, \cite{afshordi}, \cite{rassat}, \cite{
reassessment}, \cite{mcewen}, \cite{pietrobon_NVSS}, \cite{
fosalba_1}, \cite{fosalba_2}, \cite{vielva}, \cite{liu}, \cite{
ho} and \cite{giannantonio}, just to name a few of them.

The standard method for detecting the cross-correlation between the
LSS and the CMB
involves comparing the observed cross-correlation
function with its theoretical prediction for a given fiducial
cosmological model. 
The theoretical prediction
is by construction an ensemble average over all possible realizations
of the universe given the fiducial parameters,
including all possible realizations of the local matter
distribution. Assuming ergodicity, this 
ensemble average can also be thought of as an average over all
possible positions of the observer in the Universe ('cosmic mean').

However, the ISW effect is created by the decay of the
gravitational potential coming from the structures on the largest scales,
i.e. from structures
that have not yet undergone significant gravitational collapse and
are still not decoupled from the expansion of the Universe. These largest
scales are most affected by cosmic variance. Therefore, when comparing the
observed (local) cross-correlation function to its cosmic mean value,
the realization of the
matter distribution in our vicinity acts as a source of systematic noise in the
estimation of the cross-correlation, hence leading to a biased
detection significance, due to cosmic variance.

In this work, we estimate the contribution of this local variance effect
to the total variance in the detected signal
under the simplifying assumption that there is no
shot noise in the galaxy distribution. 
We find that the local variance in the detected signal
amounts to 11 per cent in the   
case of an ideal LSS survey going out to about redshift 2 and covering
enough volume
to include the large scales relevant for the ISW. 
This
agrees with \cite{cabre}, who compare different methods to
estimate the error in standard cross-correlation studies. They find
that their MC1 error, which ignores the variance coming from the
realization of the matter field and only considers the variance in
the CMB fluctuations, systematically underestimates the
error by about 10 per cent.

From the above-mentioned surveys, the local matter distribution is known
to a certain degree, and hence the local variance effect can be
reduced by working conditional on that information. 
We present a generic technique of how to include the
knowledge of the matter distribution into the detection of the ISW
via cross-correlation, hence reducing the sources of noise
to the unknown part of the matter distribution and the primordial
CMB fluctuations. We define the systematic noise that
comes from the known part of the matter distribution as bias, for it
can be removed by working conditional on the LSS data.
Our method is referred to as optimal method,
in contrast to the standard method for ISW detection mentioned above.
The main idea of the optimal method is to create an ISW template from a Wiener
filter reconstruction of the LSS. We
then use this template to
detect the amplitude of the ISW template rather than of the
theoretical cross-correlation function. 
This makes the variance in the estimated amplitude, and hence the
signal-to-noise ratio both depend on the actual realization of the matter in the
Universe. 
For an ideal LSS survey, we show that the average variance in
the detected amplitude is reduced by 13 per cent in the optimal
method.
Rephrased in terms of the signal-to-noise ratio, this reduction of the
noise leads on average to a higher detection 
significance by about 7 per cent. 
As we show in this work, in the framework of the optimal method it is
straightforward to correct for the 
magnification bias due to gravitational lensing, as described by
\cite{LoVerde}.
Furthermore, there
is no need to estimate the covariance matrix by Monte Carlo
simulations as in the standard method. This saves time and increases
the accuracy of the method.

To our knowledge, the only suggested methods for ISW detection
besides this work that also do not suffer from the local variance
effect are by \cite{zhang} and \cite{carlos}. The former
involves combining lensing-LSS cross-correlation
measurements with the ISW-LSS cross-correlation, and thereby relies on
the nowadays still-difficult lensing measurements
\citep{hu_lensing,hu_okamoto}.  
The latter follows an approach very similar to ours.
However, \cite{carlos} does not use a Wiener filter reconstruction of the LSS
distribution but works directly on the sphere. In contrast to our work
he neglects the shot noise,
which could be easily included in his analysis, though. 
In this work we will go one step further than \cite{carlos} and derive
the correct way of including the information encoded in the ISW for
cosmological parameter estimation.

Many of the above-mentioned cross-correlation studies have attempted
to constrain cosmological parameters using a likelihood function
for the cosmological parameters $p$
given the observed
cross-correlation function. Just like the detection significance, these
parameter estimates are biased by local variance. 
Furthermore, to our knowledge, there is no straightforward way of 
combining the likelihood function for the 
cross-correlation with the likelihoods for CMB  and LSS data.
In this work, we derive the
correct joint likelihood function $P(T,\delta_g \,|\, p)$ for
cosmological parameters, given the
CMB map $T$ and the LSS data $\delta_g$, from first
principles for the linear LSS formation regime. This joint likelihood
consistently includes the coupling between the two datasets introduced
by the ISW effect, which so far has been neglected in analyses
deriving cosmological parameter constraints by combining CMB 
and LSS data \citep{sdss_wmap,wmap_3}.

The article is organized as follows. We start by briefly
describing the integrated Sachs-Wolfe effect in section \ref{isw}, and
explain in detail the different
stochastic processes that are relevant for our analysis and the
correction for the magnification bias in section \ref{stochastic}. 
In section \ref{standard} we review the
standard method for detecting the ISW via cross-correlation and
estimate the contribution of the local variance to the total
variance of the detected signal. Section \ref{optimal_method} is
devoted to presenting the 
optimal method of ISW detection we developed, and to comparing it in
detail to the 
standard method. 
We discuss the role of the biasing effect due to
local variance in parameter constraints and derive the
joint likelihood function $P(T,\delta_g \,|\, p)$ in section \ref{params}.
Concluding remarks on our work are given in section
\ref{conclusions}.

\section{The Integrated Sachs-Wolfe effect} \label{isw}

The effect of decaying gravitational potential fluctuations on the CMB
is called the integrated Sachs-Wolfe effect and is described by
\be \label{isw_int}
T_\ISW(\vec{\hat n}) = 2\int_{\eta_\ls}^{\eta_0} \Psi'\left(\eta, (\eta_0 -
\eta)\, \vec{\hat n}\right)\, d\eta,
\ee
where $\eta$ denotes the conformal time, $\eta_\ls$ and $\eta_0$
the conformal time at last scattering and the present epoch, respectively, and
$\vec{\hat n}$ is the direction on the sky. Note
that the integral in the above equation has to be taken along the
backwards light cone. $\Psi$
is the gauge invariant Bardeen potential \citep{bardeen}, which
coincides with the Newtonian gravitational potential in the
Newtonian gauge used in this work.
The prime denotes the derivative with respect to
conformal time.
In order to keep the notation simple we have redefined $T_\ISW \equiv
(T_\ISW - T_0)/T_0$, where $T_0 = 
2.725\,\rm{K}$ is the temperature of the CMB monopole. We will use this
convention as 
well for $T$ and $T_\prim$, which will be defined in section
\ref{stochastic}. 

In Newtonian gauge, $T_\ISW$ is obtained by applying a linear
operator $\hm$ to the present matter density contrast $\delta_m(\eta_0)$:
\be\label{hammuMona}
T_\ISW = \hm \delta_m(\eta_0).
\ee
The matter density
contrast is defined as
$\delta_m(\vec{x}) \equiv
\left[\rho_m(\vec{x}) - \bar \rho_m\right]/ \bar \rho_m$,
where
$\rho_m(\vec{x})$ 
denotes the 
density of matter in the Universe at position $\vec{x}$, and $\bar
\rho_m$ is the background matter density.
Eq. (\ref{hammuMona}) can be verified by using the perturbation
equations derived by e.g. \cite{ks} or \cite{durrer}.

In order to obtain the expression for the operator $\hm$ in the
subhorizon-limit, let us look at the
Poisson equation  
\be
\Delta \Psi = \frac{3 H_0^2}{2} \, (1+z)\, \Omega_m \delta_m,
\ee
where $H_0$ is the present value of the Hubble constant, $\Omega_m \equiv
\rho_{m,0}/\rho_{\crit,0}$ the present ratio of matter density to
critical density, $z$ denotes the redshift, and $\Delta$ is
the Laplace operator in comoving coordinates. From the Poisson
equation we obtain
\be \label{kernel}
\Psi'(\vec{k},\eta) = \frac{3H_0^2 \Omega_m}{2 k^2} \, H(\eta)\,
\left(1-f(\eta)\right) \, \growth(\eta) \, 
\delta_m(\vec{k},\eta_0), 
\ee
where $k$ stands for the absolute value of $\vec{k}$, $H(\eta)$ is the Hubble
constant at conformal time $\eta$, $f\equiv d\ln 
\delta_m/d\ln a$ is the growth function, $\growth(\eta) \equiv
\delta_m(\vec{k},\eta)/\delta_m(\vec{k},\eta_0)$ denotes the linear
growth factor, 
and we define  
Fourier transformed quantities by
\be
\delta_m(\vec{k},0) = \int_V \!\! d^3x \, e^{i\vec{k}\bm{\cdot}\vec{x}} \, \delta_m(\vec{x},0). 
\ee
The expression for the operator $\hm$ can then be obtained by Fourier
transforming eq. (\ref{kernel}) and inserting it into eq.
(\ref{isw_int}). Note, though, that we have not used the
subhorizon-limit in this work, for eq. (\ref{hammuMona}) is valid
on superhorizon-scales as well.
\footnote{The correct formula for $\hm$
in Newtonian gauge, which also holds on superhorizon-scales, can be
obtained by differentiating and Fourier transforming
\bea \nonumber
\Psi(\vec{k},\eta) &=& \exp\left( -\int_0^\eta p\,(\vec{k},\eta')
\,d\eta' \right) 
\int_0^\eta 
\frac{H_0^2 \, \Omega_m}{2 H a^2}  \\ \nonumber
&&\times \growth(\vec{k},\eta') \,\delta_m(\vec{k},\eta_0)
\exp\left( \int_0^{\eta'} p(\vec{k},\eta'') \,d\eta''
\right) d\eta',
\eea
and inserting it into eq. (\ref{isw_int}), instead of the expression
for $\Psi(\vec{k},\eta)$ in the subhorizon-limit, eq. (\ref{kernel}).
Here, we have defined 
$p(\vec{k},\eta) \equiv \frac{k^2 + 3a^2H^2}{3aH}$
and the linear growth factor $\growth(\vec{k},\eta) \equiv
\frac{\delta(\vec{k},\eta)}{\delta(\vec{k},\eta_0)}$, which in general
depends on the Fourier mode $\vec{k}$.
}

\section{Stochastic processes} \label{stochastic}
\subsection{Realization of the matter distribution} \label{delta_m}

In the standard cosmology adopted in this work, there are different
stochastic processes to be considered. For simplifying the notation,
let us define  
\be
\g(\chi,C) \equiv \frac{1}{\sqrt{|2\pi C|}}
\exp \left(-\frac{1}{2}\chi^\dagger\,C^{-1}\chi \right)
\ee
to denote the probability density function 
of a Gaussian distributed vector $\chi$ with zero mean, given
the cosmological parameters $p$ and the covariance matrix $C \equiv \langle
\chi \chi^\dagger \rangle$, where the averages are taken over the
Gaussian distribution $\g(\chi ,C)$. Note that in general the covariance
matrix depends on the cosmological parameters, which is not
explicitly stated in our notation.
A daggered vector or matrix denotes its transposed and complex
conjugated version, as usual. 
Hence, given two vectors $a$ and $b$, $a\, b^\dagger$ must be
read as the tensor product,
whereas $a^\dagger \,b$ denotes the scalar product. Note that
these conventions can still be used for vectors and matrices in
function-spaces, like, e.g., the matter overdensity field
$\delta_m$, which 
is a continuous function of the position $\vec{x}$.

During inflation, the matter
density perturbations have been created from quantum fluctuations.
This stochastic process was close to Gaussian
\citep{mukhanov}, permitting to write down the probability
distribution for the matter density contrast given the cosmological
parameters $p$ as
\be \label{prior}
P(\delta_m \,|\, p) = \g(\delta_m ,S),
\ee
where the covariance matrix
$S \equiv \langle\delta_m \delta_m^\dagger\rangle_{P(\delta_m \,|\,
p)}$,
depends on the cosmological parameters $p$.
The average $\langle..\rangle_{P(\delta_m \,|\, p)}$ is defined as ensemble
average over the 
different realizations of $\delta_m$, the index
${P(\delta_m \,|\, p)}$ explicitly states which probability
distribution the average has to be taken over.
Given homogeneity and isotropy, we note that the Fourier
transformation of $S$ is diagonal 
\be
\langle \delta_m(\vec{k}) \delta_m(\vec{k'})^*\rangle_{P(\delta_m
  \,|\ p)} = (2\pi)^3 \delta(\vec{k}-\vec{k'}) P(k),
\ee
where $P(k)$ is the power spectrum, $\delta(..)$ denotes the Dirac
delta function, and the star is used for denoting complex conjugation.

The stochastic process due to the inflationary quantum fluctuations
created the angular fluctuations in the CMB, 
that is, the primordial fluctuations originating from the surface of last
scattering at redshift $z=1100$, as well as the integrated Sachs-Wolfe effect
imprinted by the more local 
matter distribution at $z < 2$.
Throughout this work we will assume that the primordial
fluctuations and the ISW are stochastically independent, which is
a safe assumption (apart from the very large scales), given that they
are associated with matter perturbations of very different wavelengths
\citep{x_ray_boughn}, so that very little intrinsic cross-correlation
can be expected.  
In fact, for notational convenience we will
use the symbol
$\delta_m$ to only denote the local matter distribution at $z<2$.
The joint probability distribution for $T_\ISW =
\hm\delta_m$ and 
the primordial temperature fluctuations $T_\prim$ then factorizes
\be
P(T_\ISW,T_\prim \,|\, p) 
= P(T_\ISW \,|\, p)\,P(T_\prim \,|\, p),
\ee
with
\be
P(T_\ISW \,|\, p) = \g(T_\ISW ,C_\ISW),
\ee
and
\be \label{T_prim}
 P(T_\prim \,|\, p) = \g(T_\prim ,C_\prim),
\ee
where we have defined the angular two-point auto-correlation function for
the fluctuation $T_X$ ($X$ being 'ISW' or 'prim')
\be
C_X \equiv \langle T_X T_X^\dagger\rangle_{P(T_X \,|\, 
p)}.  
\ee
Again, given homogeneity and isotropy,
$C_X$ is diagonal in spherical harmonics space 
\be \label{C_l}
\langle a_{lm}^{X} a_{l'm'}^{X\,*}\rangle_{P(T_X \,|\, p)} = C_l^X
\delta_{ll'}\,\delta_{mm'},
\ee
where $C_l^X$ is the angular power spectrum of the quantity $X$, and
we have used the expansion
coefficients of $T_X$ into spherical harmonics $Y_{lm}$,
\be \label{a_lm}
a_{lm}^X \equiv \int_S d\Omega \,T_X(\vec{\hat n})\, Y_{lm}^*(\vec{\hat n}),
\ee
where the integral is taken over the sphere.
Given that the joint distribution $P(T_\ISW,T_\prim \,|\, p)$
factorizes into two Gaussian distributions, the sum $T =
T_\ISW + T_\prim$, which denotes the temperature fluctuation of the
CMB, is again 
Gaussian distributed 
\be
P(T \,|\, p) = \g(T ,C_\CMB),
\ee
with
\be \label{sum}
C_\CMB = C_\ISW + C_\prim.
\ee
Given the cosmological parameters, the angular power spectra
$C_l^\CMB$, $C_l^\ISW$ and 
$C_l^\prim$  can all be calculated
using CMBFAST (\texttt{http://ascl.net/cmbfast.html}, \cite{cmbfast}), CAMB
(\texttt{http://camb.info}, \cite{camb}),
or CMBEASY (\texttt{www.cmbeasy.org}, \cite{cmbeasy}). In particular,  
$C_\ISW$ can be obtained from the three-dimensional matter covariance matrix
$S$ by 
\be \label{HSH}
C_\ISW = \hm S \hm^\dagger,
\ee
where we have used that linear transformations of Gaussian
random variables are again Gaussian distributed, with the covariance matrix
transformed accordingly \citep[see also][]{cooray}. 

\subsection{CMB detector noise} \label{det}

From CMB detectors, we do not read off the real $T$ as defined in
the last section, but a temperature where the detector noise $T_\Det$ has
been added.
Again this can be modeled as a Gaussian random process,
\be \label{T_det}
P(T_\Det) = \g(T_\Det, C_\Det),
\ee
where $C_\Det$ denotes the detector noise covariance. This process is
independent of the process that created the real (noiseless) $T$,
such that if we 
redefine $T \equiv T + T_\Det$ to be the temperature
we read off our detector, we obtain
\be
P(T \,|\, p) = \g(T, C_\CMB + C_\Det),
\ee
with $C_\CMB$ being the covariance matrix of the real (noiseless) CMB.

However, in most of this work we will neglect the detector
noise, for the ISW is only present on the largest angular scales,
where the dominant source of noise is cosmic variance \citep{afshordi_manual}.
The only part where we include the detector noise will be
in section \ref{params}, where we derive the joint likelihood for the
cosmological parameters, given CMB  and LSS data, for in this
likelihood we also include smaller angular scales.

\subsection{Shot noise}

Unfortunately, the matter distribution is not directly known, and we have to
rely on LSS catalogues from which we can try to reconstruct it.
A process to be considered when working with
such catalogues is the stochastic distribution of the galaxies, which only
on average follows the
matter distribution. Since the galaxies are discrete sources from which
we want to infer the properties of the underlying matter overdensity
field, we have to deal with shot noise in the galaxy
distribution. More specifically, we assume the observed number
$N_g(\vec{x_i})$ of galaxies in a volume 
element $\Delta V(\vec{x_i})$ at a discrete position $\vec{x_i}$ to be
distributed according to a Poisson distribution
\be \label{poisson_orig}
P(N_g(\vec{x_i}) \,|\, \lambda(\vec{x_i})) =
\frac{\lambda(\vec{x_i})^{N_g(\vec{x_i})} 
e^{-\lambda(\vec{x_i})}}{N_g(\vec{x_i})!}. 
\ee
Here, $\lambda(\vec{x})$ denotes the expected mean number of observed
galaxies within $\Delta V(\vec{x})$, given the matter density contrast,
\be
\lambda(\vec{x}) = w(\vec{x}) \,\overline{n_g^r} \, \Delta V \,\left[1 +
b\, \delta_m(\vec{x})\right]. 
\ee
In the above equation, $\overline{n_g^r} \equiv N_g^{r,\,\tot}/V$
denotes the cosmic mean galaxy 
density, with $N_g^{r,\,\tot}$ being the total number of galaxies in the
volume $V$. Note that we have added an index '$r$' to stress that
these are the actual (real) number of galaxies present in $\Delta
V$, not the observed number 
of galaxies $N_g$, which can be smaller due to observational detection
limits. The window
$w(\vec{x}) \equiv \Phi(\vec{x}) \,m(\vec{\hat n})$ denotes the
combined selection function $\Phi(\vec{x})$ and sky mask $m(\vec{\hat
  n})$ of the survey, and $b$ the galaxy bias, 
which in general depends on redshift, scale and galaxy type.
The variance in the observed number of galaxies $N_g(\vec{x})$ within $\Delta
V(\vec{x})$ is then $\sigma_g^2(\vec{x}) \equiv \langle 
\left[N_g(\vec{x}) - \lambda(\vec{x})\right]^2 \rangle_{N_g} =
\lambda(\vec{x})$. Here, we have used the 
index $N_g$ on the average to indicate the average over the Poisson
distribution in eq. (\ref{poisson_orig}).

If the average number of galaxies $\lambda(\vec{x})$ is large, the Poisson
distribution is well approximated by a Gaussian distribution around
$\lambda(\vec{x})$. For simplicity
we will use the Gaussian approximation throughout this work. 
Furthermore we
will ignore the dependence of the noise on $\delta_m(\vec{x})$ by using
$\sigma_g^2(\vec{x}) = 
w(\vec{x})\,\overline{n_g^r}\, \Delta V$ instead of the correct noise term
$\sigma_g^2(\vec{x}) = 
\lambda(\vec{x})$, for the latter would require a non-linear and
iterative approach. Such
an approach is beyond the scope of this paper, but is also irrelevant for the
main finding of this work. However, see 
Kitaura et al. (in preparation) and \cite{ift} for
a better handling of the Poisson noise and bias variations. 

Since the cosmic mean galaxy density $\overline{n_g^r}$ is not known,
we have to estimate 
it from the observed galaxy counts by
\be
\overline{n_g^r} \,\Delta V \equiv \frac{N_g^\tot}{\sum_{i=0}^{N_\pix}
w(\vec{x_i})}, 
\ee
where $N_g^\tot$ is the total number of observed galaxies and the sum
goes over all the pixels in our volume.

With the above-mentioned simplifications, we can now work with the
following linear data model.
First we define the the observed 
galaxy density contrast at position $\vec{x}$ to be 
\be \label{gal_overd}
\delta_g(\vec{x}) \equiv \frac{N_g(\vec{x}) -
w(\vec{x})\, \overline{n_g^r}\, \Delta V}{\overline{n_g^r}\, \Delta V},
\ee  
which is the convention used in Kitaura et al. (in
preparation). Note that this definition differs from the one usually
used in cross-correlation studies by a factor of $w(\vec{x})$ 
\citep[see e.g.][]{pogosian}.
We then write
\be \label{model_paco}
\delta_g = R\,\delta_m + \epsilon,
\ee
where $\epsilon(\vec{x})$ is 
the additive noise-term that originates in the Poissonian distribution
of $N_g(\vec{x})$, 
and $R$ is the linear response operator. In the simplest case,
$R(\vec{x_i},\vec{x_j}) \equiv b\,w(\vec{x_i}) \,\delta_{ij}$, but in
general $R$ maps the 
the continuous space in which $\delta_m$ lives onto the discrete pixel
space of our data $\delta_g$,
and it can also include the mapping from redshift-space onto comoving
coordinate space.
 
Gravitational lensing introduces a magnification
bias in the observed galaxy density contrast, as described by \cite{LoVerde}.
In our data model, it is straightforward to take this effect into
account by letting 
\bea \nonumber
R\, \delta_m (\vec{\hat n},z) \!\!\! &\equiv& \!\!\! w(\vec{\hat n}, z)
[ b\,\delta_m(r(z)\,\vec{\hat n},z) 
+ 3\,\Omega_m \,H_0^2 \left(2.5\,s(z)-1\right) \\ \nonumber
&& \!\!\!  \times \int dz' \frac{1}{H(z')}
  \frac{r(z')(r(z)-r(z'))}{r(z)} \\
&& \;\;\;\;(1+z')\, \delta_m(r(z)\,\vec{\hat n},z')],
\eea
where $r(z)$ is the comoving distance corresponding to redshift $z$, and
the slope $s$ of the number count of the source galaxies is
defined as 
\be
s \equiv \frac{d \log_{10}N(<m)}{dm},
\ee
with $m$ being the limiting magnitude and $N(<m)$ being the count of
objects brighter than $m$.
Note that in order to get the correct formula for the magnification
bias term in 3 dimensions, we used the Dirac delta function as the
normalized selection function used by \citeauthor{LoVerde}, $W(z,z')
\equiv \delta(z-z')$.

From the Poisson distribution in eq.
(\ref{poisson_orig}), we see that
$\langle \delta_g \rangle_{N_g} = R\,\delta_m$,
and hence with the above simplifications the noise $\epsilon$
is Gaussian distributed around zero 
\be \label{poisson_noise}
P(\epsilon \,|\, p) = \g(\epsilon ,N),
\ee
with the noise covariance matrix
\be
N(\vec{x_i},\vec{x_j}) \equiv \langle \epsilon(\vec{x_i})\,
\epsilon(\vec{x_j}) \rangle_{N_g} 
= \frac{w(\vec{x_i})}{\overline{n_g^r}\, \Delta V} \, \,\delta_{ij}.
\ee

\section{Standard cross-correlation method}\label{standard}
\subsection{Description}\label{decription}

In this section, we briefly review the standard method for
detecting the cross-correlation of the CMB with the projected galaxy
density contrast, which was first described by \cite{x_ray_boughn}, but see
for example also \cite{ho} and \cite{giannantonio}).
Note that we use the word galaxy density contrast
for convenience, but the method is of course the same when working with
other tracers of the LSS as mentioned in section \ref{intro}.

The theoretical cross-correlation function of two quantities
$X(\vec{\hat n})$ and 
$Y(\vec{\hat n})$ on the sky is defined in spherical harmonics space as
\be
C_l^{X,Y} \equiv \langle a_{lm}^{X}
a_{lm}^{Y\,*}\rangle_\all. 
\ee
The average in the above definition is an ensemble average over all possible
realizations of the universe 
with given cosmological parameters, i.e. over 
$P(\delta_m,\delta_g,T \,|\, p)$.  This is indicated
by the index 'all' on the average.  
We will denote the abstract cross-correlation function as a vector in
Hilbert space by $\xi^{X,Y}$ to simplify the notation. This can be understood
as a vector in pixel space or as a vector in $a_{lm}$-space. Only when
evaluating the expressions we derive, we will choose the representation of the
abstract vector $\xi^{X,Y}$ in spherical harmonics space,
$(\xi^{X,Y})_{lml'm'} = C_l^{X,Y}\delta_{ll'}\delta_{mm'}$.
In the following we will work with the cross-correlation function of the
projected galaxy density contrast with the CMB temperature fluctuations,
$\xi^{g,\,\CMB}$, in order to reproduce the standard approach in the
literature.  

The observed projected galaxy density contrast
$\delta_g^\proj$ for a redshift bin centered around redshift $z_i$
in a given direction $\vec{\hat n}$ on the sky is
\bea \nonumber
\delta_g^\proj(\vec{\hat n},z_i) \!\!\! &=& \!\!\!\!\! \int dz\, W(z,z_i)
\,\delta_g(\vec{\hat n},z)\\
\!\!\! &=& \!\!\!\!\!  \!\! \int\! dz W(z,z_i) \! \left[R
   \,\delta_m(\vec{\hat n},z) + 
\epsilon (\vec{\hat n}, z)\right]\!,\label{gal_kernel}
\eea
where $W(z,z_i)$ denotes the normalized selection function that
defines the $i$th bin, and
$\delta_g$ is given by eq. (\ref{gal_overd}). 
Note that in many cross-correlation studies the normalized selection
function $\Phi(\vec{x})$ of the survey is used to define the
bin. However, since later on we will consider a perfect galaxy survey
covering all the redshift range relevant for the ISW, we need to
introduce the additional narrow selection function $W(z,z_i)$ defining the bin.

If the LSS survey and the CMB map cover the full sky, it is convenient to
define 
an estimator for the cross-correlation function of the projected
galaxy density contrast with the CMB in
spherical harmonics space \citep{rassat},
\be \label{cc_est}
\widehat C_{l}^{g,\CMB} \equiv \frac{1}{2l+1}\sum_m Re\left(a_{lm}^{g}
a_{lm}^{\CMB\,*}\right),
\ee
where $a_{lm}^g$ and $a_{lm}^\CMB$ are the expansion coefficients of
the observed
$\delta_g^\proj$ and $T$ into spherical harmonics as defined in eq.
(\ref{a_lm}). The hat has been added to discriminate the
estimator of the 
cross-correlation function from its theoretical counterpart $C_l^{g,\CMB}$.
In the case that the experiments cover only a part of the sky, one has
to take into account the effects of mode-coupling when working in spherical
harmonics space. In this case it is therefore more straightforward to
define other
estimators for the cross-correlation function, such as averages over
the sphere in real space \citep[see e.g.][]{giannantonio} or
quadratic estimators as in \cite{afshordi}. 
However, for the statement we will make in this work the
actual definition of the estimator is not relevant, and we find
the one defined in spherical harmonics space the most convenient to
work with, since a 
closely related quantity also appears within the framework of the
optimal detection method presented later on in section \ref{optimal_method}. 
Again we use the abstract notation
$\widehat \xi^{g,\CMB}$ for the estimator of the cross-correlation
$\xi^{g\,\CMB}$. 
In order to keep the notation simple, we will from now on
understand $\widehat 
\xi^{g,\CMB}$ and $\xi^{g\,\CMB}$ as being vectors in spherical
harmonics-space as well as in bin-space, containing the
cross-correlation functions for all the different bins.

In the literature, the probability distribution of the above-defined
estimator $\widehat \xi^{g,\CMB}$ around the theoretical
cross-correlation function 
$\xi^{g,\CMB}$ is usually approximated by a Gaussian,
\be \nonumber \label{like_cc}
P\left(\widehat \xi^{g,\CMB} \,|\, p\right) =
G\left(\widehat \xi^{g,\CMB} - \xi^{g,\CMB},C \right),
\ee
where the covariance matrix of the cross-correlation estimator is defined as 
\be
C \equiv \langle\left(\widehat \xi^{g,\CMB} \!\!-\!
\langle\widehat \xi^{g,\CMB}\rangle_\all \right)\!\! \left(\widehat
\xi^{g,\CMB} \!\!-\! 
\langle\widehat \xi^{g,\CMB}\rangle_\all \right)^\dagger\rangle_\all.
\ee

The first question usually addressed in the above-mentioned
cross-correlation studies is whether a non-zero cross-correlation
function can be detected at all. 
To this end one assumes a fiducial cosmological model, which is used
to predict the theoretical cross-correlation function and covariance
matrix $C$. In this work we use the flat $\Lambda$CDM model with 
parameter values given by \cite{wmap_5}, table 1: $\Omega_b h^2 =
0.02265, \Omega_\Lambda = 0.721,\, h = 0.701,\, n_s = 0.96,\, \tau = 0.084,\,
\sigma_8 = 0.817$."
The covariance matrix is usually estimated by Monte Carlo
simulations (see \cite{cabre} for an overview), or analytically as in
\cite{afshordi}. 
The analytical prediction is possible in
the case that the joint probability distribution for
the projected galaxy density contrast
and CMB given the cosmological parameters,
$P(\delta_{gi}^\proj,\delta_{gj}^\proj,T_{\CMB} \,|\, p)$, is
Gaussian, which is valid in the framework of linear perturbation
theory. Here we have used the index $gi$ to denote the projected
galaxy 
density contrast of bin $i$.
Then the covariance matrix in spherical harmonics space can be
expressed in terms of two-point correlation functions as 
\be \label{cov_cc}
C_l(i,j) =\! \frac{1}{(2l+1)f_{\rm sky}} \!\left[
  C_l^{gi,\CMB}C_l^{gj,\CMB} \!+\! C_l^{gi,gj} \,C_l^{\CMB} \right]\!,
\ee
where we have used the auto-correlation power spectrum for the
CMB, as defined in eq. (\ref{C_l}). $C_l^{gi,gj}$
contains by definition the power coming from the underlying matter
distribution plus the shot noise. Note that, in principle, $C_l^\CMB$
in the above 
formula also includes detector noise, which we neglect here
as discussed in section \ref{det}. 
$f_{\rm
sky}$ is the fraction of the sky covered by both, the galaxy survey
and the CMB experiment. In the following we will assume $f_{\rm
sky} = 1$. 

Putting an
amplitude or fudge factor $A_\cc$ in front of the theoretical
cross-correlation function $\xi^{g,\CMB}$ by hand, one can
now find out whether it is possible to detect a non-zero $A_\cc$.
The index 'cc' on the
amplitude indicates that it is the amplitude of the cross-correlation
function. Of course this amplitude should be one in the fiducial
model. However, even if the data are taken from a universe in which
the underlying cosmology is the fiducial model we will in
general not estimate the amplitude to be one. This is due to the
different sources of stochastical uncertainty or noise in the estimate
of $A_\cc$, which we have described at length in section \ref{stochastic}.
The likelihood for the amplitude given the cosmological parameters 
reads
\be
P\left(\widehat \xi^{g,\CMB} \,|\, A_\cc,p\right) =
G\left(\widehat \xi^{g,\CMB} - A_\cc\, \xi^{g,\CMB}, C\right).
\ee
A commonly used estimator of the amplitude $A_\cc$ is the maximum
likelihood amplitude 
\bea \nonumber  
\widehat A_\cc &=& \frac{
\xi^{g,\CMB\,\dagger} \; C^{-1} \; \widehat \xi^{g,\CMB} }{
\xi^{g,\CMB\,\dagger} \; C^{-1} \; \xi^{g,\CMB}} \\ 
&=& 
\frac{\sum_l (2l+1) \sum_{i,j} C_l^{gi,\CMB} C_l^{-1}(i,j)
    \widehat C_l^{gj,\CMB}} 
{\sum_l (2l+1) \sum_{i,j} C_l^{gi,\CMB} C_l^{-1}(i,j)
  C_l^{gj,\CMB}}, \label{A_cc_def}  
\eea
where in the second line we have used the representation of the
cross-correlation 
functions in spherical harmonics space.
The maximum likelihood amplitude is an unbiased estimator (if the
underlying probability distribution is Gaussian), hence for the
fiducial model we have for the average over all cosmic realizations
\be
\langle \widehat A_\cc\rangle_\all = 1,
\ee
since $\langle \widehat C_{l}^{gi,\,\CMB}\rangle_\all = C_l^{gi,\, \CMB}$ by
definition of the latter quantity.
Note that here we have assumed that the data are taken
in a universe where the underlying cosmology is actually the fiducial
model. This will be assumed in the rest of the paper as well.

The variance in $\widehat A_\cc$ is given by
\bea \nonumber 
\sigma_\cc^2 \!\!\! &\equiv& \!\!\!\!\!\!
\langle\left(\widehat A_\cc-\langle \widehat A_\cc\rangle_\all
\right)^2\rangle_\all \\ \nonumber 
&=& \!\!\!\!\!\! \left(\xi^{g,\CMB\,\dagger} \; C^{-1} \; \xi^{g,\CMB}\right)^{-1}
\\
&=& \!\!\!\!\!\!
\left[\sum_l (2l+1) \!\sum_{i,j}\! C_l^{gi,\CMB} C_l^{-1}\!(i,j)  C_l^{gj,\CMB}
\right]^{-1}\!\!. \label{N_cc_def}
\eea

In the standard literature, an estimated significance is given to
the detection of the amplitude, the estimated signal-to-noise ratio
\bea \nonumber
\left( \frac{\widehat S}{N}\right)_\cc \!\!\!\!\!\!\!\! &\equiv& \!\!\!\!\!
\frac{\widehat 
A_\cc}{\sigma_\cc}  \\
&=& \!\!\!\!\!\!\! \frac{\sum_l (2l+1) \sum_{i,j} C_l^{gi,\CMB} C_l^{-1}(i,j)\,
    \widehat C_l^{gj,\CMB}} 
{\sqrt{\sum_l (2l+1) \sum_{i,j} C_l^{gi,\CMB} C_l^{-1}(i,j)
    \, C_l^{gj,\CMB}}}.\label{S/N_cc} 
\eea
However, since the real signal is $A_\cc = 1$, the actual signal-to-noise
ratio is given by
\bea \label{S/N_cc_real} 
\left( \frac{S}{N}\right)_\cc \!\!\!\!\!\! &\equiv& \!\!\!\!\!
\frac{1}{\sigma_\cc} \\ \nonumber
&=& \!\!\!\!\!
\sqrt{\sum_l (2l+1)\! \sum_{i,j} C_l^{gi,\CMB} C_l^{-1}(i,j)
    C_l^{gj,\CMB}},
\eea 
and is therefore independent of the data.

\subsection{Analysis of error-contributions}\label{errors}

In this section we
analyse the different sources of noise that contribute to the total variance
in eq. (\ref{N_cc_def}). 
In order to simplify this task we assume that there is no
shot noise in the galaxy distribution, that is, we set $\epsilon = 0$ in
eq. 
(\ref{model_paco}), which means that the galaxies trace the matter distribution
perfectly. 
Furthermore we work with the ideal case that we have a galaxy survey that
covers the whole sky and goes out to redshift 2.
With these two assumptions we have a perfect knowledge of the
matter distribution $\delta_m$ relevant for the ISW effect.

For sufficiently narrow bins, the integration kernels for ISW and galaxy
density contrast are approximately constant over the bin and hence
$a_{lm}^{\ISW(i)} = {\rm const(i)} \times a_{lm}^{gi}$. In eqs
(\ref{A_cc_def}), (\ref{N_cc_def}), and (\ref{S/N_cc_real}), we can
therefore substitute every index $gi$ by the index ISW($i$),
for the constant factor cancels out. Now, if one uses the ISW kernel, 
working with several narrow bins that cover the
whole volume relevant for the ISW effect, is equivalent to working
with only one bin covering the same volume. This is because the ISW
integrated over the whole relevant volume is exactly the information
about the ISW contained in the CMB. Hence one does not gain anything by
working with bins if using the correct kernel. We outline the proof
for that in Appendix \ref{bins}. 
In what follows, we therefore consider only one bin, which
significantly simplifies the form of
eqs (\ref{A_cc_def}), (\ref{N_cc_def}), and
(\ref{S/N_cc_real}).

\begin{figure}
 \centering
 \includegraphics{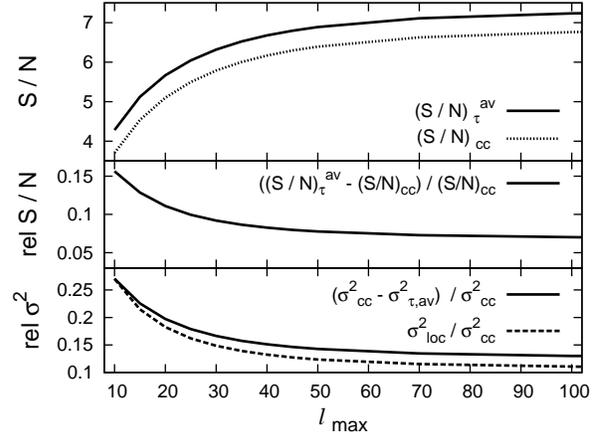}
 \caption{Comparison of the average signal-to-noise ratio and variance of the
 optimal method with the ones of the standard method for $z_\Max =
 2$. Top panel: 
 Average signal-to-noise ratio of the optimal method (solid) and
 signal-to-noise ratio of the standard method (dashed) versus the maximal
 multipole considered in the analysis. Middle panel: Relative
 improvement of the average signal-to-noise ratio in the optimal
 method. Bottom panel: Average relative
 improvement of the variance in the optimal method (solid) and relative
 contribution of the local variance to 
 the total variance in the standard method (dashed)}
 \label{compareSToN}
\end{figure}

Furthermore we note that, since the ISW is
uncorrelated with the primordial CMB fluctuations, we have
$C_l^{\ISW,\CMB} = C_l^\ISW$.
The index 'all' now indicates an average
over the probability distribution 
$P(T_\ISW,T_\prim \,|\, p) = P(T_\ISW \,|\, p) P(T_\prim \,|\,
p)$ (cf. section \ref{stochastic}).
Under the above assumptions, eq. (\ref{A_cc_def}) for the estimated
amplitude reads 
\be \label{A_cc_simple}
\widehat A_\cc = \frac{\sum_l (2l+1)
\frac{\widehat C_{l}^{\ISW,\CMB}}{C_l^{\ISW} + 
C_l^{\CMB}} }{\sum_l (2l+1)
\frac{C_l^{\ISW}}{C_l^{\ISW} + C_l^{\CMB}}},
\ee
with the variance (eq. \ref{N_cc_def})
\be \label{sigma_cc_simple}
\sigma_\cc^2 = \left(\sum_l (2l+1)
\frac{C_l^{\ISW}}{C_l^{\ISW} + C_l^{\CMB}}\right)^{-1},
\ee
and the signal-to-noise ratio in eq. (\ref{S/N_cc_real})
simplifies to
\be \label{S/N_cc_real_simple}
\left( \frac{S}{N}\right)_\cc = \sqrt{\sum_l (2l+1)
\frac{C_l^{\ISW}}{C_l^{\ISW} + C_l^{\CMB}}}.
\ee
The signal-to-noise ratio as a function of the maximum
summation index $l_\Max$ for our fiducial model is depicted in the top
panel of Fig. \ref{compareSToN}, for which we have modified CMBEASY
in order to obtain
$C_l^\ISW$ and $C_l^\CMB$.
There are 
contributions to the signal-to-noise up to roughly $l=100$. Note, though,
that our assumptions of Gaussianity of the matter realization
$\delta_m$ and the assumption of $\hm$ being a linear operator do not
hold on small scales where structure growth has become
non-linear. However, this issue will not be addressed in this work and it
will not affect our main results, which are due to advantages of our method on
the 
very large scales, which are most affected by cosmic variance.

The above estimator for the amplitude is only unbiased when averaging
over the joint distribution
\bea
\langle \widehat A_\cc\rangle_\all \equiv \langle\langle
\widehat A_\cc\rangle_\prim\rangle_{\ISW} = 1. 
\eea
Here we indicate averages over $P(T_\prim \,|\, p)$ and $P(T_\ISW
\,|\, p)$ by the indices 'prim' and '$\ISW$', respectively.
This means that both the primordial CMB fluctuations and the
realization of the local matter distribution are included in the error
budget. We call the latter the local variance, indicating that it
originates in the realization of the matter distribution in our
observed Universe.
Let us now estimate the contribution of the local variance
to the total variance of 
$\widehat A_\cc$. To this end we split the variance in eq.
(\ref{sigma_cc_simple}) into two parts
\bea \nonumber 
\sigma_\cc^2 &\equiv& \langle\langle\left(\widehat A_\cc - 1
\right)^2\rangle_\prim\rangle_{\ISW} \\ \nonumber
&=& \langle\langle\left( \widehat A_\cc - \langle \widehat A_\cc\rangle_\prim
\right)^2\rangle_\prim\rangle_{\ISW} \\ \nonumber
&& + 
\langle\left( \langle \widehat A_\cc\rangle_\prim - 1
\right)^2\rangle_{\ISW} \\ \label{sigma_ISW}
&\equiv& \sigma_\prim^2 + \sigma_{\loc}^2,
\eea
where we have defined the contributions to the variance coming from
primordial CMB fluctuations, and the local variance as
$\sigma_\prim^2$ and $\sigma_{\loc}^2$, respectively. 
Both can be easily calculated, and the second contribution
turns out to be
\be
\sigma_{\loc}^2 = 2\, \frac{\sum_l(2l+1)
\frac{\left(C_l^\ISW\right)^2}{\left(C_l^\CMB +
C_l^\ISW\right)^2}}{\left( \sum_l (2l+1) \frac{C_l^\ISW}{C_l^\CMB +
C_l^\ISW}\right)^2}.  
\ee
In the bottom panel of Fig. \ref{compareSToN}, we plot the relative
contribution of 
the local to the total variance $\sigma^2_\loc/\sigma^2_\cc$ against
the maximum $l$ that we consider in the analysis, for our fiducial
cosmological model. For a maximum multipole $l_\Max=100$, this relative
contribution amounts to 
\be
\frac{\sigma_{\loc}^2}{\sigma_\cc^2} \approx 11\%.
\ee
This estimate agrees with \cite{cabre}, who compare different
error estimates for the standard cross-correlation method. They
compare what they call the
MC1 method, which only takes into account the
variance in the CMB and ignores the variance in the
galaxy overdensity, with the MC2 method, which includes also the variance
in the galaxy overdensity. Both methods rely on performing Monte Carlo (MC)
simulations 
of the CMB, and of the galaxy overdensity in the case of MC2, and the
simulations used to compare the different error estimates have
converged with an accuracy of about 5 per cent, as stated in the paper.
The result is that, compared to the MC2 method, the MC1 method
underestimates the error by about 10 per cent, which agrees well with our
estimate.

\section{Optimal method} \label{optimal_method}

Since
the expected ISW effect is known from our galaxy survey, it is
possible to find a cross-correlation estimator that does not include
the local variance
in the error-budget, but is unbiased already when averaging
conditional on the observed galaxy density contrast $\delta_g$.
We will introduce such an estimator in this section.
To this end, we first derive the posterior distribution $P(T
\,|\, \delta_g,p)$ for the CMB temperature $T$, given
the galaxy data $\delta_g$ and the cosmological parameters $p$. From that we obtain the maximum likelihood estimator
for the amplitude of the part of $T_\ISW$ which is known from the
galaxy survey. Since we assume everything to be Gaussian distributed, this
maximum likelihood estimator is equivalent to the 
estimator we obtain from an optimal matched filter approach.

Note that a different attempt to make the detection of the ISW
unbiased by the realization of the 
local matter distribution was done by \cite{zhang}. It
involves comparing CMB-galaxy and lensing-galaxy cross-correlation
functions, and hence relies on nowadays still-difficult lensing measurements.

Another work which does not suffer from local variance is by \cite{carlos}.
He implements an optimal matched filter in spherical harmonics space, and
finds by numerical comparison
that it always performs better than or equally well as the standard
method. However, \cite{carlos} works directly on the sphere, without using a
Wiener filter reconstruction of the LSS, and therefore is slightly
suboptimal in exploiting the available three-dimensional information
on galaxy positions. 

In this work, we go one step further than \cite{carlos} and derive the joint
likelihood for cosmological parameters, given CMB  and LSS data, which
includes the small coupling of the two datasets introduced by the ISW
effect (cf. section \ref{params}).

\subsection{Derivation of the posterior distribution}

Let us first ask the question what the observed galaxy density contrast
tells us about the matter distribution $\delta_m$. Given the
data model in eq. (\ref{model_paco}) and the noise distribution
in eq. (\ref{poisson_noise}), we know that
\bea \nonumber
P(\delta_g \,|\, \delta_m,p) &=& P(\delta_g - R\,\delta_m \,|\, \delta_m,p) \\
&=&
\g(\delta_g - R\,\delta_m , N).
\eea
Using the probability distribution for $\delta_m$ given in eq.
(\ref{prior}) as a prior for $\delta_m$, we obtain the joint
probability distribution $P(\delta_g,\delta_m \,|\, p) = P(\delta_g
\,|\, \delta_m,p)\, P(\delta_m \,|\, p)$ of galaxies and matter distribution
\bea \nonumber 
P(\delta_g,\delta_m \,|\, p) &=& \g(\delta_g - R\,\delta_m, N)\,
\g(\delta_m ,S)\\ \label{joint}
&=& \g(\delta_m - Dj, D)\,\g(\delta_g , RSR^\dagger + N),
\eea
where we have defined the matter distribution
uncertainty covariance matrix
\be \label{D}
D \equiv \left(R^\dagger N^{-1}R + S^{-1}\right)^{-1}
\ee
and the response over noise weighted galaxy overdensity
\be \label{j}
j \equiv R^\dagger N^{-1}\delta_g,
\ee
which can be interpreted as the information source of our
LSS knowledge \citep{ift}.
A detailed derivation of the expression for the joint probability
distribution in
eq. (\ref{joint}) can be found in Appendix \ref{app_joint}.
This distribution can be trivially integrated over $\delta_m$ in order
to obtain the evidence
\be \label{evidence}
P(\delta_g \,|\, p) = \g(\delta_g , RSR^\dagger + N).
\ee 
Therefore the posterior distribution $P(\delta_m\,|\,\delta_g,p) =
P(\delta_m,\delta_g\,|\,p)/P(\delta_g\,|\,p)$ reads 
\be \label{posterior}
P(\delta_m \,|\, \delta_g,p) = \g(\delta_m - Dj,D).
\ee
From this posterior one can directly read off the maximum a-posteriori
estimator for the matter distribution $\delta_m$
\be \label{rec}
\delta_m^\rec \equiv Dj = (R^\dagger N^{-1}R + S^{-1})^{-1}R^\dagger
N^{-1}\delta_g. 
\ee
This is the Wiener filter applied to the galaxy-overdensity
\citep{wiener_filter,zaroubi_1,zaroubi_2,fisher,erd,paco}. 
We call
this estimator a \textit{reconstruction} of the matter distribution from the
galaxy survey, hence the symbol $\delta_m^\rec$. 

With this knowledge of the matter distribution, let us now find the
posterior distribution for $T = T_\ISW + T_\prim + T_\Det$, given the
observed galaxy density contrast $\delta_g$.
The probability distribution for $T_\ISW$, obtained from the one
for $\delta_m$, eq. (\ref{posterior}), (note that we again
use that linear
transformations of Gaussian distributed 
random vectors are again Gaussian distributed; cf. eq.
(\ref{HSH})), reads
\be \label{isw_given_gal}
P(T_\ISW \,|\, \delta_g,p) = 
\g(T_\ISW - \tau, \hm D \hm^\dagger),
\ee
where we have defined the ISW template 
\be \label{tau}
\tau \equiv \hm \, \delta_m^\rec.
\ee

Since the uncertainty in the reconstructed matter distribution is not
related to the 
primordial CMB fluctuations (cf. section \ref{delta_m}),
the joint probability
distribution for $T_\ISW$, $T_\prim$, and $T_\Det$ given
$\delta_g$ factorizes: 
\bea \nonumber
P(T_\ISW, T_\prim, T_\Det \,|\, \delta_g,p)\!\! &=& \!\!\!P(T_\ISW
\,|\, \delta_g,p) 
\, P(T_\prim \,|\, p) \\ 
\!\! && \!\!\!P(T_\Det \,|\, p).
\eea
Note that in the above equation we have used the fact that the primordial
CMB fluctuations do not 
depend on the galaxy distribution. 
We now again use the fact that the sum of
stochastically independent Gaussian distributed random variables is
again Gaussian distributed with the
sum of the covariance matrices. We then obtain the posterior distribution
for $T$, given the LSS data:
\be \label{post}
P(T \,|\, \delta_g,p) 
= \g(T - \tau, \tilde C).
\ee
Here we have used the probability distributions for $T_\prim$ and $T_\Det$,
eqs (\ref{T_prim}) and (\ref{T_det}), and
we have defined the covariance matrix for the total noise
\be \label{c_tilde}
\tilde C \equiv \hm D \hm^\dagger + C_\prim + C_\Det.
\ee

As in section \ref{standard}, we will neglect the detector noise
in the rest of this 
section and only include it when deriving the likelihood in section
\ref{params}. However, if needed it can easily be included into the
following equations by replacing $C_\prim \rightarrow C_\prim + C_\Det$.

\subsection{Estimation of the ISW amplitude}

We can now ask the same question as before, namely if it is at all
possible to detect a non-zero amplitude $A_\tau$ that we put in front
of our ISW template in eq. (\ref{post}). Again we can write down the
likelihood function for the amplitude
\be
P(T \,|\, A_\tau, \delta_g, p) = 
\g(T - A_\tau \tau, \tilde C),
\ee 
and estimate the amplitude by a maximum likelihood estimator
\be \label{A_tau}
\widehat A_\tau =  \frac{T_{\CMB}^\dagger  \tilde C^{-1}
  \tau}{\tau^\dagger \tilde C^{-1} \tau} 
= \frac{\sum_{l} (2l+1) \frac{\widehat C_{l}^{\tau,
\CMB}}{\tilde C_l}}{\sum_{l} (2l+1) \frac{
\widehat C_{l}^{\tau}}{\tilde C_l}},
\ee
where we have defined the estimator $\widehat C_{l}^{\tau}$ of the ISW
auto-correlation function 
analogous to the cross-correlation estimator in eq. (\ref{cc_est}):
\be
\widehat C_{l}^{\tau} \equiv \frac{1}{2l+1} \sum_m |a_{lm}^{\tau}|^2.
\ee
This maximum likelihood amplitude is again an unbiased estimator, but
now with respect to the probability distribution conditional on $\delta_g$,
\be
\langle \widehat A_\tau\rangle_{\rm cond}=1,
\ee
where the index 'cond' on the average denotes an average over the
distribution $P(T_{\CMB} \,|\, A_\tau,\delta_g,p)$.

In other words, we have eliminated the noise component coming
from the realization of the known part of $\delta_m$, thus
reducing the sources of noise to the unknown part of $\delta_m$ and 
the primordial CMB fluctuations.
The variance in $\widehat A_\tau$ is
\bea \nonumber 
\sigma_\tau^2 &\equiv& \langle \left(\widehat A_\tau-\langle
\widehat A_\tau\rangle_{\rm cond} 
\right)^2\rangle_{\rm cond}\\ \label{sigma_tau}  
&=& 
\left(\tau^\dagger \tilde C^{-1} \tau\right)^{-1}
= \left(\sum_{l} (2l+1) \frac{
\widehat C_{l}^{\tau}}{\tilde C_l}\right)^{-1},
\eea
and we obtain the signal-to-noise ratio
\be \label{S_N_tau_real}
\left(\frac{S}{N}\right)_\tau \equiv \frac{1}{\sigma_\tau} =
\sqrt{\sum_{l} (2l+1) \frac{ 
\widehat C_{l}^{\tau}}{\tilde C_l}}.
\ee
Note that the error estimate (and hence the signal-to-noise ratio) of the
optimal method depends on the concrete LSS 
realization, and how well it is suited to detect the ISW effect. In a
universe, where by chance the local LSS does/does not permit a good ISW
detection, the error is small/large, as it should be.

We would like to point out that in the optimal method there is no need to
estimate the covariance matrices from Monte Carlo simulations, since
for a given set of cosmological parameters, the matter covariance
matrix (power spectrum) $S$ can be calculated analytically using
the fitting formula provided by \cite{bardeen_powerfit}, since it is
still linear on the scales we are interested in. $C_\prim$ can be
obtained from Boltzmann codes such as CMBEASY, and the noise
covariance $N$ can be estimated from the data.

\subsection{Comparison of signal-to-noise ratios and biasing}\label{compare}

In order to compare our method to the standard one,
let us now again make the simplifying assumption that there is no shot noise
in the galaxy distribution, and that we have a perfect galaxy survey,
as we did in section \ref{errors}.
At the end of this section, we will also approximately look at the
effects of a galaxy survey that is incomplete in redshift,
i.e. that goes out to a maximal redshift $z_\Max < 2$. 
For the perfect survey, the
shot noise covariance matrix $N$ is zero, and hence the posterior for
$\delta_m$ in eq. (\ref{posterior}) is 
infinitely sharply peaked around the reconstruction
$\delta_m^\rec$ (eq. \ref{rec}), which turns into
\bea \nonumber
\delta_m^\rec &=& (R^\dagger N^{-1}R)^{-1}R^\dagger N^{-1}\delta_g \\
&=& R^{-1} \delta_g.
\eea
Here, $R^{-1}$ should be read as the pseudo-inverse
of $R$, e.g. as defined in terms of Singular Value Decomposition (see
\cite{press} and \cite{zaroubi_1}). 

The posterior
for $\delta_m$ in eq. (\ref{posterior}) is therefore now 
a Dirac delta function 
\be
P(\delta_m \,|\, \delta_g,p) = \delta(\delta_m - R^{-1}\delta_g),
\ee
which makes our ISW template exact, and the noise covariance matrix due
to the error in the reconstruction is zero, $\hm D \hm^\dagger = 0$,
hence leaving us with $\tilde C = C_\prim = C_\CMB -
C_\ISW$. Since our perfect LSS survey covers the complete volume relevant for
the ISW, our template is now equal to the ISW-temperature fluctuations, $\tau =
T_\ISW$. We can 
then substitute all indices $\tau$ in eqs
(\ref{A_tau})-(\ref{S_N_tau_real}) 
by the index ISW, and the estimated amplitude becomes 
\be \label{A_tau_simple}
\widehat A_\tau =  \frac{\sum_{l} (2l+1) \frac{\widehat C_{l}^{\ISW,
\CMB}}{C_l^{\CMB} - C_l^{\ISW}}}{\sum_{l} (2l+1) \frac{
\widehat C_{l}^{\ISW}}{C_l^{\CMB} - C_l^{\ISW}}},
\ee
with the variance
\be \label{sigma_tau_simple}
\sigma_\tau^2 =  \left(\sum_{l} (2l+1) \frac{
\widehat C_{l}^{\ISW}}{C_l^{\CMB} - C_l^{\ISW}}\right)^{-1},
\ee
and the signal-to-noise ratio
\be \label{S/N_tau_real_simple}
\left(\frac{S}{N}\right)_\tau
=  \sqrt{\sum_{l} (2l+1) \frac{
\widehat C_{l}^{\ISW}}{C_l^{\CMB} - C_l^{\ISW}}}.
\ee

\begin{figure}
 \centering
 \includegraphics{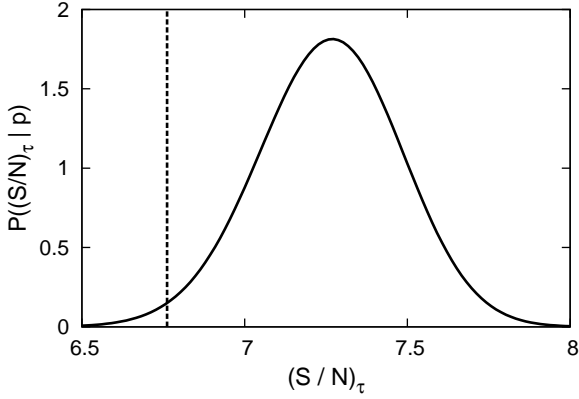}
 \caption{Probability distribution of the signal-to-noise ratio in the
 optimal method (solid) for $l_\Max=100$ and $z_\Max=2$. The vertical
 line (dashed) shows the signal-to-noise ratio of the standard method
 for comparison.}
 \label{distrSToN}
\end{figure} 
As we mentioned before, the variance, and hence the signal-to-noise ratio of
the optimal method, depend on the actual realization of the matter
distribution in our observed Universe. In Fig. \ref{distrSToN}, we plot the
probability distribution of our signal-to-noise ratio for $l_\Max =
100$ and $z_\Max = 2$, which
we have inferred from the distribution of $T_\ISW$ using the central limit
theorem for $\left(S/N\right)_\tau^2$, and from
that deriving the distribution for $\left(S/N\right)_\tau$. We have
also checked the validity of the central limit theorem in 
this case by comparing with the correct probability distribution of
the signal-to-noise ratio given by an expansion into Laguerre
polynomials as derived e.g. in \cite{pochhammer}.
The probability distribution is such that the signal-to-noise ratio
can easily differ by $\Delta \left( S/N \right)_\tau \approx 1$
for two different realizations of the matter distribution.

The mean signal-to-noise ratio $ \left(S/N\right)_\tau^\av \equiv
1/\sqrt{\sigma^2_{\tau,\,\rm{av}}} \equiv
1/\sqrt{\langle 
\sigma^2_\tau \rangle_\ISW}$ 
increases with $l_\Max$, as it
did for the standard 
method. For every $l_\Max$ we compare the mean signal-to-noise
ratio of the optimal
method to the signal-to-noise ratio of the standard method (cf. eq.
\ref{S/N_cc_real_simple}) in the top panel of Fig.
\ref{compareSToN}, again for $z_\Max = 2$. 
Note that in our formula for the signal-to-noise ratio, eq.
(\ref{S/N_tau_real_simple}), 
there is now a minus sign between $C_l^\CMB$ and $C_l^\ISW$, in contrast to
the signal-to-noise ratio of the standard method in eq.
(\ref{S/N_cc_real_simple}), which has a plus sign instead. Thus we
take advantage of 
the LSS instead of moving it into the error budget. The absolute enhancement
of the signal-to-noise ratio in our method is therefore independent of
$l_\Max$, for the main advantage of working conditional on the LSS
arises on the very large scales, where the contribution of the ISW 
to the CMB is highest.
\begin{figure}
\centering 
 \includegraphics{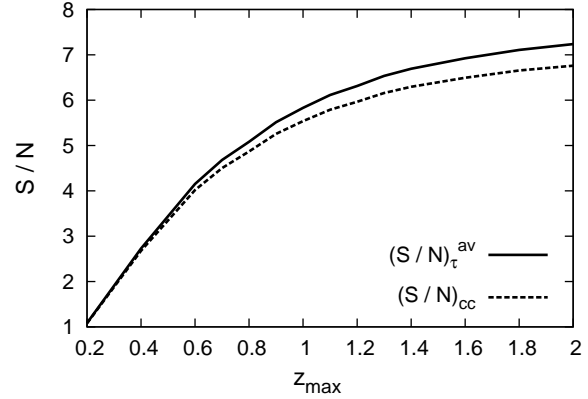}
 \caption{Average signal-to-noise ratio of the optimal method (solid) and
 signal-to-noise ratio of the standard method (dashed) versus $z_\Max$ for
 $l_\Max = 100$.}
 \label{zPlot}
\end{figure}
The average relative improvement of the signal-to-noise is depicted in
the middle panel of Fig.
\ref{compareSToN}. It amounts to about 7 per cent for $l_\Max = 100$. 
In the bottom panel of Fig. \ref{compareSToN}, we compare the mean
relative improvement  
$(\sigma^2_\cc - \sigma^2_{\tau,\,\rm{av}})/\sigma^2_\cc$ of the
variance in the optimal method with the contribution of the local to the total
variance in the standard method. The variance is reduced by about 13
per cent
in the optimal method, as compared to the standard method.

Note that the maximal average signal-to-noise
ratio we can hope for when trying to detect the ISW via
cross-correlation, given a perfect LSS survey, is $\left(S/N\right)_\tau^\av
\approx 7.3$, with a variance as depicted in Fig.
\ref{distrSToN}. Hence, if we are lucky and live in an environment
that allows for 
a high signal-to-noise ratio, we can maximally obtain a
detection significance of about $(7.5-8) \sigma$.

Let us now look at the effect of an incomplete
galaxy survey. 
Incomplete galaxy surveys
can be treated generically with our method, because the dark matter field, and
hence the ISW effect, are split into a known part (the reconstruction)
and an unknown part (an additive noise term uncorrelated with the
reconstruction).
However, for now we only want to give a rough estimate of the
consequences of an incomplete survey. Therefore we introduce a sharp
cut-off in redshift, $z_\Max$, and we simply redefine $T_\ISW \equiv
T_\ISW(< z_\Max)$ to be the part of the ISW effect created
at $z < z_\Max$. The part of the ISW that has been created at $z >
z_\Max$ is then considered part of the primordial temperature
fluctuations $T_\prim$. The power-spectra $C_l^\ISW$ and $C_l^\prim$
are redefined accordingly.
With this redefinition we have introduced
a correlation between what we consider the ISW and primordial
fluctuations, which we would not have
if we had used the reconstruction for redefining $T_\ISW$.
However, for getting the picture, we ignore this subtlety for the moment.

In Fig. \ref{zPlot}, we plot the signal-to-noise ratio of the standard
method together with the average signal-to-noise ratio of the optimal
method versus $z_\Max$ for $l_\Max = 100$, where we have used the
above-described 
redefinition of $C_l^\ISW$ in eqs (\ref{S/N_tau_real_simple}) and
(\ref{S/N_cc_real_simple}). 
With decreasing maximal redshift of the LSS survey, the total
signal-to-noise ratio in both methods goes down,  
as does its relative
enhancement of the optimal method as compared to the standard one.
Also the relative contribution of the local to the total variance in
the standard method goes down with decreasing survey depth.

As we stated in section \ref{errors}, the amplitude-estimate of the standard
method is biased when the averaging is performed conditional on the
galaxy-data $\delta_g$.
This leads to an over- or underestimation of the detection
significance, for the estimated amplitude is used
when estimating the signal-to-noise ratio from the data. As we have
shown, the contribution of the local to the total variance of the
estimator is quite small, about 11 per cent for an ideal galaxy survey and
even smaller for a shallower survey. However, we could be
unlucky and live in an unlikely realization of the matter
distribution, given the power spectrum, which would enhance the effect of
the biasing.

With the method we presented in this work, the local variance effect
is reduced. If we knew the
local matter distribution perfectly, we would not be affected by local variance
at all, as we have shown. Unfortunately, we
have to rely on reconstructions 
of the matter
distribution from LSS surveys, which suffer from shot noise, and the effects
of mask and
selection function. However, the reconstruction treats mask and
selection function in an optimal way, and extracts the maximum amount
of information from the LSS data which can then be used in the
ISW detection.

\section{Likelihood function for cosmological parameters}\label{params}

The above-described biasing effect is of course also present when
moving from the pure 
detection of the ISW to the task of constraining cosmological
parameters using the ISW, which has been attempted in many of the
above-mentioned cross-correlation studies. 
This problem can already be seen in the likelihood function
for the cosmological parameters of the standard method in eq.
(\ref{like_cc}). The 
estimator of the cross-correlation function $\widehat \xi_{g, \, \CMB}$ could
be 
quite different from the theoretical prediction with the underlying
parameter values, just because we are living in an unlikely
realization of the matter distribution, given the power spectrum. Then
the likelihood in eq. (\ref{like_cc}) would favour cosmological
parameter values for which the theoretical prediction of the
cross-correlation function is closer to its estimator, hence biasing 
the parameter estimates.

Furthermore, to our knowledge, there is no straightforward way of
combining the likelihood from 
the cross-correlation in eq. (\ref{like_cc}) with the likelihoods
for CMB  and LSS data, as e.g. given by \cite{wmap_1},
\cite{likelihood_2df_1} and \cite{likelihood_2df_2}. 
Usually, when combining CMB  with LSS data for
deriving constraints on cosmological parameters, it is assumed that
the two datasets are stochastically independent, i.e. that
$P(T,\delta_g \,|\, p) = P(T \,|\, p) \, P(\delta_g \,|\,
p)$ [cf. \cite{sdss_wmap}, \cite{wmap_3} and \cite{wmap_5}]. But the ISW
effect (and also other effects as e.g. the Sunyaev-Zel'dovich-effect)
introduces a small stochastical dependence of the CMB data on the
LSS data. That is, instead of assuming that the joint likelihood
factorizes, one should consider
\be
P(T,\delta_g \,|\, p) = P(T \,|\, \delta_g,p)\, P(\delta_g\,|\,p),
\ee
in which we insert eq.s (\ref{post})
and (\ref{evidence}), obtaining
\bea \nonumber
P\left(T,\delta_g \,|\, p\right)
&=& \g(T - \tau,\tilde C) \,\, \g(\delta_g ,RSR^\dagger +
N), \\ \label{likelihood}
&=& P(T \,|\, p)  \, P(\delta_g \,|\, p) \,Q(T, \delta_g \,|\, p),
\eea
and we recall for convenience the definition  of $\tilde C$,
eq. (\ref{c_tilde}), 
\bea \nonumber
\tilde C \equiv \hm D \hm^\dagger + C_\prim + C_\Det,
\eea
of $D$, eq. (\ref{D}),
\bea \nonumber 
D \equiv \left(R^\dagger N^{-1}R + S^{-1}\right)^{-1},
\eea
and of $\tau$, eq. (\ref{tau}),
\bea \nonumber
\tau \equiv \hm \delta_m^\rec.
\eea
In the last step in eq. (\ref{likelihood}), we have expressed the
joint likelihood in terms of the likelihoods $P(T \,|\, p)$ and
$P(\delta_g \,|\, p)$ for only CMB  and only LSS data, respectively,
and the coupling term
\bea \nonumber 
Q(T, \delta_g \,|\, p) &\equiv& \frac{P(T \,|\, \delta_g, p)}{P(T
\,|\, p)} \\
&=&
\frac{\g(T-\tau, \tilde C)}{\g(T,C_\CMB)}.
\eea

Eq. (\ref{likelihood}) is the generic expression for the joint likelihood
$P(T,\,\delta_g \,|\, p)$ for the cosmological
parameters $p$, given CMB  and LSS data, consistently including the
small coupling term $Q(T,\delta_g \,|\, p)$ between the two datasets
introduced by the ISW effect. 
The quantities depending on the cosmological parameters are 
$S,\, C_\prim,\, \hm,\, R$ and, in general, $N$. Multiplying the
likelihood by 
a prior $P(p)$ for the cosmological parameters, one can then sample
the parameter space and derive constraints on the cosmological
parameters from the posterior distribution $P(p \,|\, T,\delta_g)
\propto P(T,\delta_g \,|\, p)\,P(p)$.

Note that this likelihood function remains valid if galaxy bias variations,
position dependent noise, and other non-linear effects of galaxy
formation are taken into account, as long as the variance of the
reconstruction $D \equiv \langle \left( \delta_m^\rec - \delta_m\right) \left(
\delta_m^\rec - \delta_m\right)^\dagger \rangle$ is estimated
consistently (see \cite{ift} for methods to treat such complications).

\section{Conclusions} \label{conclusions}

Due to the obscuration by primordial CMB fluctuations, the detection
of the integrated Sachs-Wolfe effect remains a challenge, and has to
be performed by cross-correlating the CMB signal with the large-scale
structure. The standard method for doing so involves comparing the
observed cross-correlation function to its theoretical
prediction, which is by construction an ensemble average over all
realizations of the primordial CMB fluctuations and matter
distributions. Hence, the realization of the matter
distribution in our Universe acts as a source of systematic noise in
the estimate of the 
cross-correlation function, an effect that we have named the
local variance.

Since the ISW is only present on the largest scales, the effect of the
local variance is quite notable, amounting to about 11 per cent of the
total variance in the standard method for an ideal LSS survey. This
leads to a biased estimated detection significance of the
cross-correlation, and 
when moving from the pure ISW detection to
parameter estimation, it also biases the parameter constraints. 
We note that even if the local variance contributes only about 11 per cent to
the total variance of the detected signal, we could be unlucky and
live in an unlikely realization of the matter distribution, given the
power spectrum. This would enhance the effect of the bias on the detection
significance and parameter constraints.

Given that information about the matter distribution can be inferred
from the LSS survey,  
the local variance can be reduced by working conditional on
this information. 
In  this work, we have presented a generic technique of how to include the
knowledge of 
the matter distribution into ISW detection in an optimal way, hence
reducing the effect of the local variance. This optimal method
requires a three-dimensional Wiener filter reconstruction of the LSS,
including an 
estimator of the full reconstruction uncertainty covariance
matrix. Note that also other reconstruction techniques that provide an
estimator of the uncertainty covariance can easily be included into
our method.
The reduction of the local variance stresses the importance to
measure and reconstruct the LSS to the highest possible accuracy, as
aimed by \cite{paco} and Kitaura et al. (in preparation). 

The conditionality on the LSS data results in a
dependence of the variance in the detected signal on the actual
realization of the matter 
distribution in the observed Universe. The average variance in the
optimal method is reduced by about 13 per cent as compared to 
the standard method, again in the case of an ideal LSS survey.
The reduction of the noise translates into an average enhancement of the
signal-to-noise or detection significance by about 7 per cent for the
optimal method. However, note that also the signal-to-noise ratio
depends on the actual realization of the matter distribution.

We would also like to point out that in the optimal method, there is no need to
estimate the covariance matrix by Monte Carlo simulations, which safes time
and increases the accuracy of the method (using 1000 Monte Carlo simulations
to estimate the standard covariance matrix of the cross-correlation function
only reaches an accuracy of about 5 per cent, as stated by \cite{cabre}).

In order to consistently include the information 
encoded in the ISW effect for deriving cosmological
parameter constraints, we have derived the joint 
likelihood $P(T,\delta_g \,|\, p)$ for the cosmological
parameters $p$, given CMB  and LSS data within the linear regime of
structure formation. If one wishes to use the ISW
effect for constraining cosmological parameters, one should 
include the additional CMB-galaxy data coupling term $Q(T,\delta_g
\,|\, p)$, which we have factored out in eq. 
(\ref{likelihood}), into the usual likelihood analysis.

\section*{acknowledgments}
The authors would like to thank Simon White, Thomas Riller, Carlos
Hernandez-Monteagudo, Jens Jasche, Andr{\'e} Waelkens, Anthony Banday, and
Georg Robbers for useful discussions and comments. Thanks also to
Pier-Stefano Corasaniti for suggesting to include the effects of the
magnification bias. We acknowledge the use of CMBEASY.
This project was partly funded by the Transregional 
Collaborative Research Centre TRR 33 - The Dark Universe

\bibliographystyle{mn2e}
\bibliography{bibl.bib}

\begin{appendix}

\section{Proof of the equivalence of the number of bins} \label{bins}

We now outline the proof that if one uses the correct kernel, i.e. the
ISW kernel rather than the kernel for the galaxy density contrast in
the analysis, the estimated amplitude $\widehat A_\cc$ and the
variance $\sigma_\cc^2$ are independent of the number of
bins chosen, provided that all bins together cover
the whole volume relevant for the ISW effect. 
The proof here is done only for the variance, but
follows the same scheme for the estimated amplitude.
The total variance $\sigma_\cc^2$ one obtains when working with
$N$ bins is given by eq. (\ref{N_cc_def}), where we have substituted
the index $gi$ by ISW($i$), following the argument of section \ref{errors}:
\be \label{var_bins}
\sigma_\cc^2 \!=\!\! \left[\sum_l (2l+1)\! \sum_{i,j} C_l^{\ISW(i),\CMB}
  C_l^{-1}(i,j)  C_l^{\ISW(j),\CMB} \right]^{-1}\!\!\!\!\!\!\!.
\ee
We then use the form of the covariance matrix given by
eq. (\ref{cov_cc}) and the following relations that only hold for the
ISW kernel: 
\bea
C_l^{\ISW(i),\CMB} &=& \sum_{j=1}^N C_l^{\ISW(i),\ISW(j)}\\
 C_l^\ISW &=&\sum_{j=1}^N C_l^{\ISW(j),\CMB}\!.
\eea
Now we choose a fixed but arbitrary number of bins $N$, invert the
covariance matrix and by inserting the above relations we obtain
\be
\sum_{i,j} C_l^{\ISW(i),\CMB} C_l^{-1}(i,j)  C_l^{\ISW(j),\CMB} 
= \frac{C_l^\ISW}{C_l^\ISW+C_l^\CMB}.
\ee
Inserting this into eq. (\ref{var_bins}), the resulting formula for
$\sigma_\cc^2$ is exactly what we obtain from one single bin covering the whole 
volume relevant for the ISW effect. We have checked this explicitly
for $N=2..5$ and it is straightforward, though timely, to also check it
for any other number of bins.

\section{Derivation of the joint probability distribution} \label{app_joint}

In this section we will derive in detail the expression for the joint
probability distribution $P(\delta_g,\,\delta_m \,|\, p)$ given in
eq. (\ref{joint}). We start with
\bea \nonumber 
P(\delta_g,\delta_m \,|\, p) &=& \g(\delta_g - R\,\delta_m,N)\,
\g(\delta_m ,S)\\ \nonumber
&=& \frac{1}{\sqrt{|2\pi N||2\pi S|}}\\ \nonumber
&& \times \exp\left(-\frac{1}{2}(\delta_g - R\,\delta_m)^\dagger  
N^{-1}(\delta_g - R\,\delta_m) \right) \\ \label{a1} 
&& \times \exp\left(-\frac{1}{2}\,\delta_m^\dagger\,
S^{-1}\delta_m \right). 
\eea
Let us first rewrite the exponent
\bea \nonumber
&& \!\!\!\! (\delta_g - R\,\delta_m)^\dagger N^{-1}(\delta_g - R\,\delta_m)
+ \delta_m^\dagger\, S^{-1}\delta_m \\ \nonumber
&=& \!\!\!\! \delta_m^\dagger \,D^{-1} \delta_m - 2\,j^\dagger\, \delta_m +
\delta_g^\dagger \,N^{-1} \,\delta_g \\ \nonumber
&=& \!\!\!\! (\delta_m - Dj)^\dagger D^{-1}(\delta_m - Dj) - j^\dagger D\,j +
\delta_g^\dagger \,N^{-1} \delta_g \\
&=& \!\!\!\! (\delta_m - Dj)^\dagger D^{-1}(\delta_m - Dj) + \delta_g^\dagger
(RSR^\dagger + N)^{-1} \delta_g,
\eea
where we have used the definitions of $D$ and $j$, eqs (\ref{D})
and (\ref{j}), in the first step, then completed the square in the
second step, and we will separately prove the last step as Lemma 1 in the next
subsection. 
After doing that we will prove that
\be
|2\pi N| |2\pi S| = |2\pi D| |2\pi (RSR^\dagger + N)|,
\ee
which we name Lemma 2, allowing us to reformulate eq.
(\ref{a1}) as  
\bea \nonumber
&=& \frac{1}{\sqrt{|2\pi D||2\pi (RSR^\dagger + N)|}}\\ \nonumber
&& \times \exp\left(-\frac{1}{2}(\delta_m - Dj)^\dagger  
D^{-1}(\delta_m - Dj) \right) \\
&& \times \exp\left(-\frac{1}{2}\,\delta_g^\dagger\,
(RSR^\dagger + N)^{-1}\delta_g \right), 
\eea
which is what we claimed in eq. (\ref{joint}).

\subsection{Lemma 1}
In this subsection we prove that
\be \label{to_prove_orig}
j^\dagger D\,j - \delta_g^\dagger\, N^{-1}\delta_g = -\delta_g^\dagger\,
(RSR^\dagger + N)^{-1} \delta_g.
\ee
In order to simplify the notation, let us introduce
\be
M \equiv R^\dagger N^{-1} R.
\ee
It can be easily seen that eq. (\ref{to_prove_orig}) is equivalent to 
\be \label{to_prove}
N^{-1}R\,(S^{-1}\!\! + \!M)^{-1}\!R^\dagger \,N^{-1} \!- \!N^{-1} =
-(RSR^\dagger + N)^{-1}
\ee
by inserting the respective expressions for $D$ and $j$.
We start with eq. (\ref{to_prove}) and transform it
into an equation which is true.
\begin{eqnarray} \nonumber 
\!\!\!\!\!\!\!\! && \!\!\!\! N^{-1}R\,(S^{-1} + M)^{-1}R^\dagger
\,N^{-1}-\!N^{-1} \!=\! -(RSR^\dagger + N)^{-1} \\ \nonumber
\!\!\!\!\!\!\!\! &\Longleftrightarrow& \!\!\!\! RSR^\dagger N^{-1}
R(S^{-1}+M)^{-1}R^\dagger 
N^{-1} \\ \nonumber
\!\!\!\!\!\!\!\! && \!\!\!\! + R(S^{-1}+M)^{-1}R^\dagger N^{-1} -
RSR^\dagger N^{-1} -1 = -1 
\\ \nonumber
\!\!\!\!\!\!\!\! &\Longleftrightarrow& \!\!\!\!
RS[M(S^{-1}+M)^{-1}+(1+MS)^{-1}-1]R^\dagger 
N^{-1} = 0 \\ \nonumber
\!\!\!\!\!\!\!\! &\Longleftrightarrow& \!\!\!\!
RS[MS(1+MS)^{-1}+(1+MS)^{-1}-1]R^\dagger N^{-1} 
= 0 \\
\!\!\!\!\!\!\!\! &\Longleftrightarrow& \!\!\!\! RS[(1+MS)(1+MS)^{-1} -
  1] R^\dagger N^{-1} = 0.  
\end{eqnarray}
This equation is true, QED

\subsection{Lemma 2}

In the following we prove that 
\be
|2\pi N| |2\pi S| = |2\pi D| |2\pi(RSR^\dagger + N)|,
\ee
which is equivalent to 
\be
|N||S| = |D| |RSR^\dagger + N|,
\ee
for the factors of $2\pi$ cancel for matrices that 
operate on the same vector space.
Let us write
\bea \nonumber
\frac{|N||S|}{|D|} &=& |N||S||D^{-1}|\\ \nonumber
&=& |N||SD^{-1}|\\ \nonumber
&=& |N||S\, (S^{-1} + R^\dagger N^{-1}R)| \\ \nonumber
&=& |N| \exp\left( \log |1 + SR^\dagger N^{-1}R| \right)\\ \nonumber 
&=& |N| \exp\left(\rm{Tr} \,\log(1+SR^\dagger N^{-1}R)   \right)\\ \nonumber
&=& |N| \exp\left(\rm{Tr} \,\log(1+RSR^\dagger N^{-1})   \right)\\ \nonumber
&=& |N| \exp\left( \log |1 + RSR^\dagger N^{-1}| \right)\\ \nonumber 
&=& |N||RSR^\dagger N^{-1}+1| \\ \nonumber
&=& |(RSR^\dagger N^{-1} + 1)N|\\
&=& |RSR^\dagger +N|.
\eea
The crucial step here was to use the cyclic invariance of the trace Tr
and to notice that this cyclic invariance still holds for the trace of
a logarithm, which can be easily verified using the Taylor expansion
of the logarithm.\newline

\end{appendix}

\label{lastpage}

\end{document}